\documentstyle[aps,prl,multicol,epsf]{revtex}

\newcommand{\be}{\begin{equation}}
\newcommand{\ee}{\end{equation}}

\newcommand{\ba}{\begin{eqnarray}}
\newcommand{\ea}{\end{eqnarray}}

\def\sg{{\sigma}}

\def\br{{\mathbf{r}}}

\begin{document}
\bibliographystyle{unsrt}
\title{Interfaces of polydisperse fluids : surface tension and adsorption properties}

\author{L. Bellier-Castella$^{1}$, H. Xu $^{1}$ and M. Baus 
$^{2}$}
\date{\today}
\maketitle

\noindent
$^{1}$ D\'epartement de Physique des Mat\'eriaux (UMR 5586 du CNRS),\\
Universit\'e Claude Bernard-Lyon1, 69622 Villeurbanne Cedex, France\\
$^{2}$ Physique des Polym\`eres, Universit\'e Libre de Bruxelles,\\
Campus Plaine, CP 223, B-1050 Brussels, Belgium\\

\vspace{2truecm}

\begin{abstract}
We consider a system of spherical colloidal particles with a 
size polydispersity and use a simple van der Waals description in order to study
 the combined effect of both the polydispersity and the spatial non-uniformity 
induced by a planar interface between a low-density fluid phase 
(enriched in small particles) and a high-density fluid phase (enriched in large 
particles). We find a strong adsorption of small particles at the interface, the 
latter being broadened with respect to the monodisperse case. We also find that
 the surface tension of the polydisperse system results from a competition 
between the tendancy of the polydispersity to lower the surface tension and its
tendancy to raise the critical-point temperature (i.e. its tendancy to favor 
phase separation) with the former tendancy winning at low temperatures and the
latter at the higher temperatures.

\end{abstract}

\begin{multicols}{2}
\narrowtext

\section{Introduction}
\label{sec1}
\vspace{1truecm}
Many of the complex fluids used in the industry or in the soft condensed matter
physics laboratory are collections of nearly identical particles which exhibit
 one or several polydispersities, i.e. properties such as the size or the shape
of these particles which are distributed whithin some interval in an almost
 continuous manner [1]. These fluids are hence continuous mixtures of
 similar particles and it is of some practical importance to know how their
 composition
 or polydispersity influences their physical properties, e.g. their phase
 behavior and rheological properties. Here we will be concerned only with the
 equilibrium phase behavior of such polydisperse fluids. The generalization of
the well-established methods for the study of phase transitions in discrete
 mixtures to continuous mixtures is a technically very demanding task which has
recently become an active area of research [2]. Most of this research
 has been limited to spatially uniform (or bulk) phases while it is our purpose
 here to extend it further to non-uniform situations involving the interface
 between two coexisting bulk phases. Such an intrinsic interface is different
from, e.g. the interface between a bulk phase and a substrate (see [3]
for some preliminary work in this direction). Indeed, in the latter case the 
interface can be characterized thermodynamically by a surface excess free-
energy defined relative to the substrate whereas in the former case it is characterized
by a surface tension (which is both a surface free-energy and a thermodynamic
 force) defined with respect to an intrinsic ``surface of tension'' [4]. We will hence be particularly interested in the influence of the
 polydispersity on this interfacial surface tension.\par
In order to keep the problem manageable we will restrict ourselves to the fluid
 phases of a system of spherical particles with a size-distribution, a situation typical for many colloidal dispersions [5]. The initial or 
parent-phase size-distribution will be assumed to be fixed once and for all by the
 production process of the colloidal particles and taken to be of the monomodal
type, i.e. peaked around a single reference species, such as is appropriate for the
 polydisperse generalization of a one-component system. When this (initial)
 parent-phase is put into appropriate thermodynamic conditions it will phase 
separate or `` fractionate'' into two (or more [6]) daughter-phases.
Since these daughter-phases differ in density and composition an interface will build up between them across which the properties of one bulk phase transform
continuously into those of the coexisting bulk phase. The properties of such a
spatially non-uniform two-phase system are most conveniently studied in two
 steps. First, one determines the two spatially uniform bulk phases which are
 able to coexist in equilibrium. Next, one determines the profiles across the
interface of those properties which are spatially varying in the two-phase
 system. For the first step we will use the results of our earlier study [7] based on the van der Waals (vdW) free-energy of a polydisperse system of spherical colloidal particles interacting via excluded volume repulsions and 
vdW-like attractions. Such a description is of course not exact but simplifies
considerably the technical problems raised by the study of phase equilibria in
polydisperse systems. For the second step we will use an earlier extension of 
the vdW free-energy to spatially non-uniform systems [8] and generalize
 it here to polydisperse systems. In order to extract the surface tension we
 will, finally, adapt to the present case a general procedure advocated
 elsewhere [9].\par
In Sec.~\ref{sec2} we introduce the vdW density functional of a spatially non-uniform polydisperse system. The density profiles across a planar interface 
are computed in Sec.~\ref{sec3} while the corresponding adsorption
 properties will be discussed in Sec.~\ref{sec4}. The pressure profiles
 across the planar
interface are determined in Sec.~\ref{sec5} while the resulting surface
 tension is presented in Sec.~\ref{sec6}. Our conclusions follow in the final Sec.~\ref{sec7}.
\section{The spatially non-uniform polydisperse system}
\label{sec2}
The equilibrium properties of spatially non-uniform systems (e.g. interfaces)
are most easily studied within density functional theory (DFT) [10].
The starting point of DFT is the variational free-energy, $A(T,[\rho],[\Phi])$
:\\

\ba
A(T,[\rho],[\Phi])=&F(T,[\rho])\nonumber\\
&+\int d\br \int d\sigma
\,\rho(\br,\sigma) \{\Phi(\br,\sigma)-\mu(\sigma)\}
\label{eq1}
\ea
where $T$ is the equilibrium temperature, $F(T,[\rho])$ the intrinsic Helmholtz
free-energy viewed as a functional (indicated as $[\rho]$) of the average local
 number density, $\rho(\br,\sigma)$, $\br$ being the position variable 
(assuming spherical particles) and $\sigma$ the (dimensionless) polydispersity 
variable, $\mu(\sigma)$ is the chemical potential of species $\sigma$ and
 $\Phi(\br,\sigma)$ the one-body external field responsible for the spatial non-uniformity of the system (the functional dependence of $A$ on $\Phi(\br,\sigma)$
 being indicated as $[\Phi]$). The equilibrium density, $\rho(\br,\sigma)$, 
corresponding to a given $\Phi(\br,\sigma)$, can then be obtained by solving
the Euler-Lagrange equation :
\be
 \frac{\delta A(T,[\rho],[\Phi])}{\delta \rho(\br,\sg)}|_{T,[\Phi]}=0
\label{eq2}
\ee 
corresponding to (\ref{eq1}), viz. :\par
\be
\mu(\sigma)=\Phi(\br,\sigma)+{\frac{\delta F(T[\rho])}{\delta \rho(\br,\sigma)}} |_{T}\, .
\label{eq3}
\ee
Eq. (\ref{eq3}) expresses the fact that in equilibrium the chemical
 potential $\mu(\sigma)$ of each species $\sigma$ has to remain constant in
 space. In the present study, $F(T,[\rho])$, will be approximated by the
 following vdW-type expression [7,8,11] :
\ba
F(T,[\rho])=k_B\,T\int d\br \int d\sg\rho(\br,\sigma)\nonumber\\
\{\ln(
\frac{\Lambda^3(\sg)\rho(\br,\sigma)}{E(\br,[\rho])})-1\}\nonumber\\
+\frac{1}{2}\int d\br \int d\sg \int d\br' \int d\sg'\,\rho(\br,\sigma)\nonumber\\
 V_A(|\br-\br'|;\sg,\sg') \rho(\br',\sg')\nonumber\\
\label{eq4}
\ea
where $k_B$ is Boltzmann's constant, $\Lambda(\sigma)$ the thermal de Broglie
 wavelength of species $\sigma$, $V_A(r;\sigma,\sigma')$ the potential of
 attraction between two particles of species $\sigma$ and $\sigma'$ a distance
$r=|\br|$ apart, while $E(\br,[\rho])$ represents the excluded volume 
correction resulting from the repulsions :
\be
E(\br,[\rho])=1-\int d\sigma v(\sigma)\rho(\br,\sigma),
\label{eq5}
\ee
$v(\sigma)=\frac{4\pi}{3}R^3(\sigma)$ being the volume of a spherical particle
 of radius $R(\sigma)$. It should be noted that, as is usual in this context 
[11], $\sigma$ is used here both as a species label and as the
 (dimensionless) polydispersity variable, $R(\sigma)/R(1)$, $R(1)$ being the
radius of the reference species $\sigma=1$. From (\ref{eq4}) we obtain for (\ref{eq3}) :
\ba
\mu(\sg)=\Phi(\br,\sigma)+k_B\,T\ln(
\frac{\Lambda^3(\sg)\rho(\br,\sigma)}{E(\br,[\rho])})\nonumber\\
+k_B\,T\frac{v(\sg)}{E(\br,[\rho])}\int d\sg'\rho(\br,\sg')\nonumber\\
+\int d\br' \int d\sg'\,V_A(|\br-\br'|;\sg,\sg')\rho(\br',\sg').
\label{eq6}
\ea
The above represents a straightforward extension of DFT to continuous mixtures
while (\ref{eq4}-\ref{eq5}) reduces to the vdW free-energy used in [7] for the
uniform polydisperse system as well as to the vdW free-energy used in [8] for the non-uniform monodisperse system. Both approximations reducing to the usual vdW free-energy for the uniform monodisperse system. The basic physics
 of the vdW approximation being, as usual, the correction of the ideal gaz
 behavior for the finite size of the particles via the excluded volume
 correction ($E$) and the inclusion of the cohesion between the particles via
 the  interparticle attractions ($V_A$), as described here respectively by the
 first and second term of (\ref{eq4}). Of course, more involved expressions of
$F(T,[\rho])$ are available but these can only add further complications to the
already fairly complex calculations required by the present combination of the 
 non-uniformity with the polydispersity of the system. Past experience has
 shown however that the present vdW approximation is able to capture the
 essence of the underlying phase behavior in a qualitatively correct manner [12]. In [7] we have studied several model-polydispersities differing
 in the $\sigma$-dependence of $v(\sigma$) and $V_A(r;\sigma,\sigma')$. It was
found there that the model based on the simple approximation :
\be
v(\sigma)=v(1) ,
\;\; V_A(r;\sigma,\sigma')=\sigma \,\sigma'\, V_A(r;1,1)
\label {eq7}
\ee
has a phase behavior which is similar to that of the more involved expressions but is
simpler to study. Henceforth we will use thus (\ref{eq4}-\ref{eq5}) together with
 (\ref{eq7}). The physical contents of (\ref{eq7}) reflects the fact (cf. [7]) that the amplitude-polydispersity of $V_A(r;\sigma,\sigma')$ dominates the
 volume-polydispersity of $v(\sigma)$.
The inclusion of the volume-polydispersity of $v(\sigma)$ will therefore
not alter qualitatively our conclusions.
\section{The planar interface}
\label{sec3}
We will consider a planar interface perpendicular to the z-axis. Translational
 invariance in the (x,y)-directions implies then, $\rho(\br,\sigma) \rightarrow
 \rho(z,\sigma)$ and $\Phi(\br,\sigma) \rightarrow \Phi(z,\sigma)$, so that the
Euler-Lagrange eq. (\ref{eq6}) can be rewritten, after separating the local
 (in $z$) and non-local contributions, as :
\ba
\mu(\sg)=\Phi(z,\sigma)+\mu_0(z,\sigma;T,[\eta])\nonumber\\
+\sg \int_{-\infty}^{\infty} dz'\, V_1(|z-z'|) \, \{\eta_1(z')-\eta_1(z)\}
\label{eq8}
\ea
with $\mu_0(z,\sigma;T,[\eta])$ a shorthand notation for :
\ba
\mu_0(z,\sigma;T,[\eta])=k_B\,T\ln
\frac{\Lambda^3(\sg)}{v(1)}+k_B\,T\ln\frac{\eta(z,\sigma)}{1-\eta_0(z)}
\nonumber\\
+k_B\,T\frac{\eta_0(z)}{1-\eta_0(z)}+\sigma \, V_0 \, \eta_1(z)
\label{eq9}
\ea
where
\ba
\eta(z,\sigma)=v(1)\, \rho(z,\sigma) ,
\;\; \eta_0(z)=\int d\sigma \, \eta(z,\sigma) ,\nonumber\\
\;\; \eta_1(z)=\int d\sigma \, \sigma \, \eta(z,\sigma) ,
\label {eq10}
\ea
are the dimensionless density and polydispersity moments, whereas :
\ba
v(1)\, V_1(|z|)=\int_{-\infty}^{\infty} dx \int_{-\infty}^{\infty} dy\, V_A(r;1,1) 
,\nonumber\\
\;\; v(1)\,V_0=\int d\br\, V_A(r;1,1) ,
\label{eq11}
\ea
with $\sigma=1$ denoting the reference particle of volume $v(1)$. The external
(symmetry breaking) field, $\Phi(z,\sigma)$, will as usual be replaced by
 boundary conditions. We thus consider eq. (\ref{eq8}) without external
 field, viz. :
\ba
\mu(\sg)=\mu_0(z,\sigma;T,[\eta])
+\sg \int_{-\infty}^{\infty} dz'\, V_1(|z-z'|) \,\nonumber\\
 \{\eta_1(z')-\eta_1(z)\}
\label{eq12}
\ea
and require that for, $z \rightarrow \pm \infty$, the solution $\eta(z,\sigma)$
of (\ref{eq12}) matches the bulk-phase densities, say $\eta_\pm (\sigma)$, or
$\eta(z=\pm \infty,\sigma)=\eta_\pm (\sigma)$. These bulk-phase densities must
 hence satisfy eq. (\ref{eq12}) for $z=\pm \infty$. Taking the limit of (\ref{eq12}) for $z \rightarrow \pm \infty$, the second
 term in its r.h.s. will vanish and we obtain :
\be
\mu(\sigma)=\mu_0(z=\pm \infty,\sigma;T,[\eta_\pm])
\label{eq13}
\ee
i.e., the chemical potential of species $\sigma$ in the non-uniform system must
 be constant and equal to the chemical potential of species $\sigma$ in the two
 bulk-phases. Indeed, evaluating the r.h.s. of (\ref{eq13}) from (\ref{eq9}) for 
$\eta(z=\pm \infty,\sigma)=\eta_\pm(\sigma)$ we obtain :
\ba
\mu_0(\pm \infty,\sigma;T,[\eta_\pm])=k_B\,T\ln
\frac{\Lambda^3(\sg)}{v(1)}+k_B\,T\ln\frac{\eta_\pm (\sigma)}{1-\eta_0^\pm}
\nonumber\\
+k_B\,T\frac{\eta_0^\pm}{1-\eta_0^\pm}+\sigma \,V_0\, \eta_1^\pm
\label{eq14}
\ea
where :
\be
\eta_0^\pm=\int  d\sigma \, \eta_\pm(\sigma),
\;\; \eta_1^\pm=\int d\sigma \, \sigma \, \eta_\pm (\sigma)
\label{eq15}
\ee
while the r.h.s. of (\ref{eq14}) represents (cf. [7]) the chemical potential of a
 uniform phase of density $\eta_\pm(\sigma)$. When the two bulk phases are in
 equilibrium, the chemical potential of the $\eta_+ (\sigma)$ phase must be
 equal to that of the $\eta_- (\sigma)$ phase, hence (\ref{eq13}) will be
 satisfied. Eq. (\ref{eq13}) allows us to elimite $\mu(\sigma)$ from
 (\ref{eq12}) and rewrite it as :
\ba
\mu_0(\pm \infty,\sg;T,[\eta_\pm])-\mu_0(z,\sigma;T,[\eta])\nonumber\\
=
\sg \int_{-\infty}^{\infty} dz' \, V_1(|z-z'|) \, \{\eta_1(z')-\eta_1(z)\}
\label{eq16}
\ea
an integral equation for $\eta(z,\sigma)$ incorporating the boundary
 conditions. On using (\ref{eq9}) and (\ref{eq14}) we can rewrite (\ref{eq16}) as :
\be
\eta(z,\sigma)=A_0^\pm (z) M^\pm (z,\sigma)
\label{eq17}
\ee
where $A_0^\pm(z)$ is a shorthand notation for :
\be
 A_0^\pm(z)=\frac{1-\eta_0(z)}{1-\eta_0^\pm}\,\exp\{\frac{1}{1-\eta_0^\pm}
-\frac{1}{1-\eta_0(z)}\}
\label{eq18}
\ee
and $M^\pm(z,\sigma)$ for :
\be
M^\pm(z,\sigma)=\eta_\pm(\sigma) \exp \,\sigma \int_{-\infty}^{\infty} dz' \, \beta \, V_1(|z-z'|) \, \{\eta_1^\pm-\eta_1(z')\}
\label{eq19}
\ee
where $\beta=1/k_B T$. Taking now the first two $\sigma$-moments of (\ref{eq17})
yields a system of two integral equations for $\eta_0(z)$ and $\eta_1(z)$, viz. :
\ba
\eta_0(z)=A_0^\pm(z) M_0^\pm(z)\nonumber\\
\eta_1(z)=A_0^\pm(z) M_1^\pm(z)
\label{eq20}
\ea
where
\be
M_0^\pm(z)=\int d\sigma M^\pm(z,\sigma),
\;\; M_1^\pm(z)=\int d\sigma \sigma M^\pm (z,\sigma).
\label{eq21}
\ee
Solving (\ref{eq20}) and substituting the result into (\ref{eq18}-\ref{eq19})
 yields finally $\eta(z,\sigma)$ via (\ref{eq17}). Note that, since the
 bulk-phase densities $\eta_\pm(\sigma)$ must correspond to the same chemical
 potential (cf. (\ref{eq13})) :
\be
\mu_0(\infty,\sigma;T,[\eta_+])=\mu_0(-\infty,\sigma;T,[\eta_-]),
\label{eq22}
\ee
the equations (\ref{eq16}-\ref{eq20}) bearing the ``+'' sign are equivalent
 (but not identical) to those bearing the ``-'' sign.
The above equations can therefore be used with either sign, the results
will be the same provided, of course, that the two bulk phases are
coexisting equilibrium phases.\par
 To proceed we must
 specify $V_A(r;1,1)$. Since only fluid phases are involved the particular form
 of $V_A(r;1,1)$ is not very important and in view of (\ref{eq11}) we will, for
simplicity, take it to be gaussian or in the notation of [7] :
\be
V_A(r;1,1)=-\epsilon(1,1)\, 8\, v(1)\, \frac{\exp (-b\, \overline{r}^2)}
{ R^3(1)\, (\frac{\pi}{b})^{3/2}}\; ;  \;\; 
\overline{r}=\frac{r}{R(1)}
\label{eq23}
\ee
where $\epsilon(1,1)$ is a reference amplitude. If, $t=k_B T/\epsilon(1,1)$, denotes
 the dimensionless temperature eq. (\ref{eq11}) yields on using (\ref{eq23}) :
\be
 \beta V_1(|z|)=-\frac{8}{t}\, \frac{\exp (-b\, \overline{z}^2)}
{R(1)\, (\frac{\pi}{b})^{1/2}},
\;\;\overline{z}=\frac{z}{R(1)}\; ;  \;\; 
 \beta V_0=-\frac{8}{t}.
\label{eq24}
\ee
Below we have used (for convergence reasons) $b=4$.\par
Finally, eq. (\ref{eq17}) also requires explicit data for $\eta_\pm(\sigma)$.
For the latter we will take the two-phase coexistence densities obtained in [7]
for the same temperature ($t$) and for an initial parent-phase density,
 $\rho_0(\sigma)=\rho_0 \, h_0(\sigma)$, of average density $\rho_0$ (or, in dimensionless
 form, $\eta_0=\rho_0 \, v(1)$) and a Schulz-Zimm size-distribution $h_0(\sigma)$ :
\be
h_0(\sigma)=\frac{\alpha^\alpha}{\Gamma(\alpha)} \sigma^{\alpha -1} exp(-\alpha \sigma)
\label{eq25}
\ee
where $\Gamma(\sigma)$ is the Euler gamma function and $I=1+(1/\alpha)$ 
the polydispersity index. Note that $1/\alpha=I-1$ is the variance of
 $h_0(\sigma)$ so that $I=1$ (or $\alpha=\infty$) corresponds to the
 monodisperse limit whereas the reference species $\sigma=1$ corresponds to
 the average value of $\sigma$ in the parent-phase. Of course, other
 size-distributions can be used but as shown in [6,7] the particular form of $h_0(
\sigma)$ has little influence as long as it remains monomodal.\par
In the present work, we have studied two polydispersities, viz. $\alpha=50$ 
($I=1.02$) and $\alpha=15$ ($I=1.07$), for several temperatures ($t$) and densities ($\eta_0$). In Fig. 1 we show three binodals of the bulk phase diagram for
 $\alpha=50$ (see also [7]). In Fig. 2 we show the size-distributions of two bulk-phases 
coexisting for $t=1$, $\eta_0=0.48$ and $\alpha=15$. The inset of Fig. 2 shows
the corresponding density distributions. It is seen there that for a range of
 $\sigma$-values ($0.3\leq \sigma \leq 0.7$) the densities of the
 ``low''-density phase actually exceed the corresponding densities of the
 ``high''-density phase. Finally, in Fig. 3 we show a variety of density profiles for the 
non-uniform two-phase system. 
It is seen that, compared to the monodisperse case, the polydispersity widens
the interfacial region.
The results of Fig. 3 have been obtained by solving
 eq. (\ref{eq20}) iteratively (e.g. by starting from a tanh-profile) whereas the results shown
in Figs. 1-2 have been obtained as explained in [7].

\section{Adsorption properties}
\label{sec4}
In macroscopic thermodynamics it is customary to replace the continuous density
 profiles, $\eta(z,\sigma)=v(1)\, \rho(z,\sigma)$, obtained in 
Sec.~\ref{sec3} by discontinuous (with respect to $z$) profiles,
 $\overline{\eta}(z,\sigma) =v(1)\, \overline{\rho}(z,\sigma)$, of the 
form [4] :
\ba
\overline{\rho}(z,\sigma)=\rho_+(\sigma)\, \theta(z-z_G(\sigma))+\rho_-(\sigma)\,\theta(z_G(\sigma)-z)\nonumber\\
+\Gamma(\sigma)\,\delta (z-z_G(\sigma))
\label{eq26}
\ea
where $\eta_\pm (\sigma)=v(1)\,\rho_\pm(\sigma)$ are the density distributions 
of the two coexisting bulk phases and $\Gamma(\sigma)$ is the adsorption of
 species $\sigma$ at the interface for which $z=z_G(\sigma)$ is the Gibbs
 dividing surface of species $\sigma$. In (\ref{eq26}), $\theta(z)$ denotes the
 Heaviside step function and $\delta(z)$ the Dirac delta function. The
 macroscopic ($\overline{\rho}(z,\sigma)$) and microscopic ($\rho(z,\sigma)$)
 profiles can be adjusted by requiring them to satisfy :
\be
\int_{-\infty}^{\infty} dz \, \{\rho(z,\sigma)-\overline{\rho}(z,\sigma)\}=0
\label{eq27}
\ee
which implies that $\Gamma(\sigma)$ be defined as the surface excess density,
 viz. :
\be
\Gamma(\sigma)=\int_{-\infty}^{\infty} dz \, \{\rho(z,\sigma)-\hat{\rho}(z,\sigma)\}
\label{eq28}
\ee
where $\hat{\rho}(z,\sigma)$ is the following bulk-phase switch function :
\be
\hat\rho(z,\sigma)=\rho_+(\sigma)\, \theta(z-z_G(\sigma)) +\rho_-(\sigma)\, \theta(z_G(\sigma)-z).
\label{eq29}
\ee
Since at $z=\pm \infty$ both densities match, $\rho(z=\pm \infty,\sigma)=\hat
\rho(z=\pm \infty,\sigma)=\rho_\pm(\sigma)$, we can integrate (\ref{eq28}) by parts and obtain :
\be
\Gamma(\sigma)=\int_{-\infty}^{\infty} dz \, (z_G(\sigma)-z)\,\rho'(z,\sigma)
\label{eq30}
\ee
where $\rho'(z,\sigma)=\partial \rho(z,\sigma)/ \partial z$. As seen from
 (\ref{eq30}) the value of $\Gamma(\sigma)$ attributed to a given
 $\rho(z,\sigma)$ still depends on the value of $z_G(\sigma)$. Since there is
 no absolute determination possible for $z_G(\sigma)$ we have to fix it
 arbitrarily, e.g. for the reference species $\sigma=1$. Taking henceforth
 $z_G(\sigma)\equiv z_G(1)$ for all $\sigma$ we can fix $z_G(1)$ by requiring that the
 corresponding adsorption, $\Gamma(1)$, vanishes. Eq. (\ref{eq30}) implies
 then : 
\be
z_G(1)=\frac{\int_{-\infty}^{\infty} dz \, z \, \rho'(z,1)}{\int_{-\infty}^{\infty} dz \, \rho'(z,1)}
\label{eq31}
\ee
whereas (\ref{eq30}) becomes :
\be
\Gamma_1(\sigma)=\int_{-\infty}^{\infty} dz \, (z_G(1)-z) \, \rho'(z,\sigma)
\label{eq32}
\ee
where the subscript $1$ on $\Gamma(\sigma)$ indicates that the adsorption of
species $\sigma$ is referred to the zero-adsorption Gibbs dividing surface
 (\ref{eq31}) of the reference species $\sigma=1$, hence $\Gamma_1(1)=0$.
Since, moreover, the system is of infinite extend in the z-direction we may
choose $z_G(1)$ as the origin of our coordinate system, i.e. $z_G(1)=0$.
Some examples of $\Gamma_1(\sigma)$ are given in Fig. 4. As seen from Fig. 4, at the interface there is both an excess of small particles ($\Gamma_1(\sigma)>0$ for $\sigma<1$) and a depletion of large particles ($\Gamma_1(\sigma)<0$ for $\sigma>1$)
with an adsorption ($\Gamma_1(\sigma)$) which strongly depends on $t$ and 
$\eta_0$. 

\section{Pressure profile across the planar interface}
\label{sec5}
Besides the density profile ($\rho(z,\sigma)$)which gives rise to the adsorption
properties described in the previous section, an interface also involves a
 pressure profile ($p(z)$) which in turn gives rise to the surface tension as 
will be shown [9] in the next section. In order to expose the pressure
 in the interior of the interface described by $\rho(z,\sigma$), we first cut
 this interface with a plane perpendicular to the density profiles, say the
 $x=0$ plane, and remove the matter on the $x<0$ side of this plane while
 leaving the matter on the $x\geq 0$ side intact. Such a semi-infinite system
with a density, $\rho(\br,\sigma)=\theta(x)\,\rho(z,\sigma)$, can be realized
within the DFT of Sec.\ref{sec2} by replacing the matter removed from the
 $x<0$ half-space by a corresponding external field, say $\Phi(\br,\sigma)$. 
The pressure acting normal to the $x=0$ plane, i.e. acting in a direction which
 is tangential to the density profiles, can then be obtained by submitting the 
$x=0$ plane to an infinitesimal non-uniform normal deformation, viz. $x\rightarrow x+\delta u(y,z)$, and computing the resulting thermodynamic work of
 deformation (cf. [9]). Since during this infinitesimal deformation,
$\Phi(\br,\sigma) \rightarrow \Phi(\br,\sigma)+\delta \Phi(\br,\sigma)$ and
$\rho(\br,\sigma) \rightarrow \rho(\br,\sigma)+\delta \rho(\br,\sigma)$, the
system has to remain in equilibrium at the given $T$ and $\mu(\sigma)$, the
relation between $\delta \Phi(\br,\sigma)$ and $\delta \rho(\br,\sigma)$ can be
 obtained from the equilibrium condition (\ref{eq3}) as :
\ba
\delta \mu(\sigma)=0=\delta \Phi(\br,\sigma)+\int d\br' \int d\sigma' \nonumber\\
\frac{\delta^2 F(T,[\rho])}{\delta \rho (\br,\sigma) \,  \delta \rho (\br',\sigma')} \, \delta \rho (\br',\sigma').
\label{eq33}
\ea
The resulting thermodynamic work of deformation, $\delta A$, can then be
 obtained from (\ref{eq1}) :
\be
\delta A |_{T,[\mu]}=\int d\br \int d\sigma \, \rho(\br,\sigma) \, \delta \Phi(\br,\sigma)
\label{eq34}
\ee
or on using (\ref{eq33}), from :
\ba
\delta A |_{T,[\mu]}=-\int d\br \int d\sigma \int d\br' \int d\sigma' \, 
\rho(\br,\sigma) \, \nonumber\\
\frac{\delta^2 F(T,[\rho])}{\delta \rho (\br,\sigma) \,
  \delta \rho (\br',\sigma')} \,
\delta \rho (\br',\sigma').
\label{eq35}
\ea
 Since in the present geometry we have, $\delta \rho(\br,\sigma)=\rho(x+\delta u
(y,z),y,z,\sigma)-\rho(x,y,z,\sigma)=\delta u(y,z) \,. (\partial \rho(\br,\sigma
)/\partial x) +\mathit{O}(\delta u^2)$, eq. (\ref{eq35}) can be rewritten after
dropping the $\mathit{O}(\delta u^2)$ term:
\be
\delta A |_{T,[\mu]}=-\int dy' \int dz' \, \delta u(y',z') \, p(y',z')
\label{eq36}
\ee
which defines the pressure $p(y,z)$ acting at
 $\br=(0,y,z)$ in a direction normal to the $x=0$ plane. Indeed, since $\delta u(y,z)$ is arbitrary
 (\ref{eq35}-\ref{eq36}) imply :
\ba
p(y,z)=\int dx \int d\sigma \int d\br' \int d\sigma' \, \rho(\br',\sigma')
\nonumber\\
\frac{\delta^2 F(T,[\rho])}{\delta \rho (\br',\sigma') \,
\delta \rho (\br,\sigma)}
.\frac{\partial \rho(\br,\sigma)}{\partial x}
\label{eq37}
\ea
where, for convenience, we have interchanged the role of the primed and 
unprimed variables. In the present vdW-approximation we obtain from (\ref{eq4})
 :
\ba
\frac{\delta^2 F(T,[\rho])}{\delta \rho(\br,\sigma)\,\delta \rho(\br',\sigma')}=k_B\,T \frac{\delta (\br -\br') \, \delta(\sigma-\sigma')}{\rho(\br,\sigma)} +
\; \nonumber\\
V_A(|\br-\br'|;\sigma,\sigma')\nonumber\\
+k_B\,T \, \frac{\delta(\br-\br')}{E(\br,[\rho])}\{v(\sigma)\nonumber\\
+v(\sigma')+\frac
{v(\sigma) \, v(\sigma')}{E(\br,[\rho])} \, \int d\sigma'' \, \rho(\br,\sigma'')
\} 
\label{eq38}
\ea
which on behalf of (\ref{eq7}), reduces here to :
\ba
\frac{\delta^2 F(T,[\rho])}{\delta \rho(\br,\sigma)\,\delta \rho(\br',\sigma')}=k_B\,T \frac{\delta (\br -\br') \, \delta(\sigma-\sigma')}{\rho(\br,\sigma)} +
\;\nonumber\\
 \sigma \sigma' \, V_A(|\br-\br'|;1,1)\nonumber\\
+k_B\,T \, \delta(\br-\br')\, v(1) \{\frac{1+E(\br,[\rho])}{(E(\br,[\rho]))^2}
\} 
\label{eq39}
\ea 
so that (\ref{eq37}) can be rewritten :
\ba
p(y,z)=\int dx \int d\sigma \{k_B\, T \frac{\partial \rho(\br,\sigma)}{\partial
 x}\nonumber\\
+ \int d\br' \int d\sigma' \, \rho(\br',\sigma') \, V_A(|\br-\br'|;\sigma,\sigma') \frac{\partial \rho(\br,\sigma)}{\partial x}\nonumber\\
+k_B \, T \, v(1) \, \rho(\br,\sigma) \frac{(1+E(\br,[\rho]))}{(E(\br,[\rho]))^2} \, \int d\sigma' \, \frac{\partial \rho(\br,\sigma')}{\partial x}\}.
\label{eq40}
\ea 
Taking into account that here, $\rho(\br,\sigma)=\theta(x)\, \rho(z,\sigma)$,
 we can rewrite (\ref{eq40}) as :
\ba
v(1)\, p(y,z)=k_B \, T \int_{-\infty}^{\infty} dx \, \frac{\partial}{\partial x}\{\frac{\eta_0(z)\, \theta (x)}{1-\eta_0(z)\,\theta(x)}\}\nonumber\\
 +\int_{-\infty}^{\infty} dx
\int d\br' \, V_A(|\br-\br'|;1,1)\, \delta (x) \, \eta_1(z) \, \theta(x') \, \eta_1(z')
\label{eq41}
\ea 
where $\eta_0(z)$ and $\eta_1(z)$ have been defined in (\ref{eq10}). Eq. (\ref{eq41}) can be rewritten as :
\be
v(1) \, p(z)=\frac{k_B T \, \eta_0(z)}{1-\eta_0(z)} + \frac{1}{2} \, \eta_1(z) 
\, \int_{-\infty}^{\infty} dz' V_1(|z-z'|) \, \eta_1(z')
\label{eq42}
\ee 
where $V_1(|z|)$ was defined in (\ref{eq11}) and we took into account that
 $p(y,z)$ is independant of $y$ as expected from the translational invariance
 in the y-direction. We finally rewrite (\ref{eq42}) in a manner similar to 
(\ref{eq8}) :
\ba
v(1) \, p(z)=v(1) \, p_0(z;T,[\eta])\nonumber\\
+ \frac{1}{2} \, \eta_1(z) 
\, \int_{-\infty}^{\infty} dz' V_1(|z-z'|) \, \{\eta_1(z')-\eta_1(z)\}
\label{eq43}
\ea 
with, $p_0(z;T,[\eta])$, a shorthand notation for :
\be
v(1) \, p_0(z;T,[\eta])=\frac{k_B T \, \eta_0(z)}{1-\eta_0(z)} + \frac{1}{2} \,V_0 \, (\eta_1(z))^2 
\label{eq44}
\ee 
where $V_0$ was defined in (\ref{eq11}). It is seen that (\ref{eq44})
 represents the usual vdW-pressure of a uniform (polydisperse) system evaluated
 for the local density $\eta(z,\sigma)$, while the same is true of (\ref{eq9})
for the chemical potential. From (\ref{eq40}) it is seen that the local
 pressure $p(z)$ is completely determined by the local density $\eta(z,\sigma)$
. From (\ref{eq43}) and $\eta(\pm \infty,\sigma)=\eta_\pm(\sigma)$ we obtain $
p(\pm \infty)=p_0(\pm \infty;T,[\eta_\pm])$, but since $\eta_\pm(\sigma)$ must
satisfy [7] :
\be
\frac{k_B T \, \eta_0^+}{1-\eta_0^+} + \frac{1}{2} \,V_0 \, (\eta_1^+)^2=\frac{k_B T \, \eta_0^-}{1-\eta_0^-} + \frac{1}{2} \,V_0 \, (\eta_1^-)^2
\label{eq45}
\ee
together with (\ref{eq22}), we have $p(\infty)=p(-\infty)$, or $\int dz \, p'(z
)=0$, which expresses the stability of the planar interface (cf. [9] for details).\par
 Some of the pressure profiles, $p(z)$, obtained from (\ref{eq43}) using the
 density profiles of Sec.~\ref{sec3} are shown in Fig. 5. While $p(z)$ remains
 constant in the bulk phases it exhibits a structure in the interfacial region
which is more pronounced for the lower temperatures and disappears gradually
 when $t=t_c$ is approached. This structure consists of a pressure depletion on
the high-density side of the interface and a pressure excess on the low-density
 side (a similar structure was found in the monodisperse case [8] for a 
different potential $V_A(r;1,1)$ and was seen also in the simulation results 
of [13]). This local structure of $p(z)$ reflects a competition between the
 characteristic length-scales of $\rho(z,\sigma)$ and of $V_A(r;1,1)$. It is 
also seen (compare Figs. 5(a) and (b)) that increasing the polydispersity (i.e.
 lowering $\alpha$) widens the interfacial region (i.e. the region where $p'(z)\ne 0$). 
 
\section{Surface tension and surface of tension}
\label{sec6}
In a way analoguous to (\ref{eq26}), the microscopic pressure profile $p(z)$ of
 Sec.~\ref{sec5} is replaced in macroscopic thermodynamics by a discontinuous
 pressure profile, $\overline{p}(z)$ :
\be
\overline{p}(z)=p_+ \, \theta(z-z_0)+p_- \, \theta(z_0-z) -\gamma \, \delta(z-z_0)
\label{eq46}  
\ee
where $p_\pm=p(\pm \infty)$ denote the bulk-phase pressures and $\gamma$ is the
 surface tension acting on a surface of tension located at $z=z_0$. As in (\ref{eq27})
the two profiles can be adjusted by requiring that:
\be
\int_{-\infty}^{\infty} dz \, \{p(z)-\overline p(z)\}=0
\label{eq47}
\ee
which on behalf of (\ref{eq46}) yields :
\be
\gamma=\int_{-\infty}^{\infty} dz \, \{\hat{p}(z)-p(z)\}
\label{eq48}
\ee
where $\hat{p}(z)$ :
\be
\hat{p}(z)=p_+ \, \theta(z-z_0)+p_- \, \theta(z_0-z)
\label{eq49}
\ee
is the switch function for the bulk-phase pressure (cf. (\ref{eq29})). 
Integrating
 (\ref{eq48}) by parts and taking into account that for a planar interface we
 must have, $p(\pm\infty)=\hat{p}(\pm \infty)=p_\pm$ together with $p_+=p_-$,
 yields:
\be
\gamma=\int_{-\infty}^{\infty} dz \, z \, p'(z)
\label{eq50}
\ee
which shows that $\gamma$ can be determined from the knowledge of $p'(z)=dp(z)/
dz$ alone, i.e. without knowing $z_0$ (cf. the difference with
 (\ref{eq30})), a feature specific to the planar interface (cf. [9]). To
 determine $z_0$ one can nevertheless also impose (cf. [9]) that :
\be
\int_{-\infty}^{\infty} dz \, z \, \{p(z)-\overline p(z)\}=0
\label{eq51}
\ee
which on using (\ref{eq47}) can be rewritten as :
\ba
0=\int_{-\infty}^{\infty} dz \, (z-z_0) \, \{p(z)-\overline p(z)\}\nonumber\\
=\int_{-\infty}^{\infty} dz \, (z-z_0) \, \{p(z)-\hat{p}(z)\}
\label{eq52}
\ea
or integrating (\ref{eq52}) by parts and using $p_+=p_-$ one obtains :
\be
\int_{-\infty}^{\infty} dz \, (z-z_0)^2 \, p'(z)=0.
\label{eq53}
\ee
Using, $\int dz \, p'(z) =0$, eq. (\ref{eq53}) yields finally :
\be
z_0=\frac{1}{2} \, \frac{\int_{-\infty}^{\infty} dz \, z^2 \, p'(z)}{\int_{-\infty}^{\infty} dz \, z \, p'(z)}
\label{eq54}
\ee
so that both $\gamma$ and $z_0$ can be obtained from $p'(z)$ (cf. (\ref{eq50}) and (\ref{eq54})).\par
Fig. 6 shows some of the results obtained for $\gamma$ using the pressure
 profiles of 
Sec.~\ref{sec5}.
From Fig. 6a it is seen that $\gamma$ decreases when increasing $\eta_0$ at a
 fixed polydispersity. When, instead, the polydispersity is changed starting
 from the monodisperse case ($\alpha=\infty$)  there are two competing effects 
(cf. Fig. 6b) : the polydispersity tends to lower the surface tension but at the
 same time it raises the critical-point temperature since the polydispersity
 favors the phase separation (cf. [7]). As a net result of this competition the surface
 tension of the polydisperse system is lower than that of the monodisperse one
 for $t<t_0$ but exceeds it for $t>t_0$ with a crossover temperature $t_0$ which 
increases with the polydispersity. To show this more clearly Fig. 6c displays 
$\gamma$ versus $t/t_c(\alpha)$, where $t_c(\alpha)$ is the critical point 
temperature of the system with polydispersity $\alpha$.\par
The above surface tension ($\gamma$) acts on the surface of tension located at
$z=z_0$ (cf. eq. (\ref{eq46})). From Fig. 7 it is seen that $z_0<0$, i.e. the surface of tension is different from the zero-adsorption Gibbs dividing surface
($z=z_G(1)=0$) and located on the high-density side of this interface (this can
 also be seen from Fig. 5). The fact that $z_0\ne z_G(1)$ points to a 
fundamental inadequacy of the macroscopic description of interfaces (cf. [9]) 
since two different quantities are used to locate the interface on a 
macroscopic level. The quantity, $l_T=z_G(1)-z_0$, is usually called Tolman's
 length [4]. Finally, from Fig. 8 it is seen that on approaching the critical
 point $\gamma$ vanishes with a classical critical exponent ($=3/2$) as 
expected from the mean-field vdW-theory [4].

\section{Conclusions}
\label{sec7}
We have studied the planar interface resulting from the phase separation or
fractionation of a parent phase of a polydisperse colloidal system into a 
low-density fluid phase enriched in small particles and a high-density fluid
phase enriched in large particles. In order to tackle the combined effect of
 the system's polydispersity and spatial non-uniformity we have kept the 
theorical description as simple as possible but are confident that similar
 results can be found from more involved descriptions. Using a simple van der Waals description [7-8] to model the polydisperse non-uniform system of 
spherical colloidal particles with excluded volume repulsions and gaussian 
attractions it was found that the small particles accumulate at the interface, 
the latter being moreover depleted with larger particles and broadened with respect to
the monodisperse case. We also found that for a given temperature the surface 
tension is the result of a competition between two polydispersity-induced 
effects, namely its tendancy to lower the surface tension and at the same time 
to raise the critical point temperature, with the former effect winning at 
low-temperatures and the latter at higher temperatures. Finally, the surface 
of tension was found to be located on the high-density side of the interface 
pointing to a positive Tolman's length.\par  

\noindent
{\bf Acknowledgements}
\newline
M.B. acknowledges financial support from the F.N.R.S.
\pagebreak
\noindent{\bf Figure Captions}\par
\vspace{1truecm}
\noindent
{\bf FIG. 1.} The temperature ($t$)-density ($\eta$) bulk-phase diagram for
 $\alpha=50$ (cf. [7]). Three binodals are shown. They correspond to the parent-phase densities $\eta_0=0.3$ (dashed line), $\eta_0=\eta_c=0.3659$ (full line) and $\eta_0=0.45$ (dot-dashed line). Two 
binodals are truncated upwards at resp. the supra-critical temperature $t\simeq
1.246$ (for $\eta_0=0.3$) and the infra-critical temperature $t\simeq 1.1905$ (for $\eta_0=0.45$) while the untruncated binodal passes through the critical
 point $t_c=1.2355$, $\eta_c=0.3659$. Similar results are obtained (cf. [7]) for
 $\alpha=15$ in which case the critical point corresponds to $t_c=1.2889$, $\eta_c=0.4842$.\par
 {\bf FIG. 2.} The density-($\eta_{\pm}(\sigma)=\eta_{\pm} \, h_{\pm}(\sigma)$; cf. inset) and size-distributions ($h_{\pm}(\sigma)$) of the low-density 
($\eta_{+}$; full line) and the high-density ($\eta_-$; dashed line) bulk phases ($\eta_+<\eta_-$) which coexist for $t=1$, $\eta_0=\eta_c=0.4842$ and
 $\alpha=15$. Note from the inset that for, $0.3\lesssim \sigma \lesssim 0.7$,
 the behaviors of $\eta_\pm(\sigma)$ and $\eta_\pm=\int d\sigma \, \eta_\pm(
\sigma)$ are reversed, i.e. although $\eta_+<\eta_-$ we have $\eta_+(\sigma)>
\eta_-(\sigma)$ for these $\sigma$-values. The low (high)-density phase is enriched in small (large) particles.\par
 {\bf FIG. 3.} Density profiles for $\alpha=15$ across a planar interface :
 (a)-(b) at constant temperature ($t=1$) and (c) at constant density ($\eta_0=\eta_
c=0.4842$). In (a) we show $\eta_0(z)$ (full line) and $\eta_1(z)$ (dotted
 line). Also shown for comparison is the monodisperse case (dashed line)
 corresponding to $\alpha=\infty$ and $\eta_0(z)\equiv \eta_1(z)$. It is seen that the polydispersity broadens the interface. In (b) we show $\eta(z,\sigma)$ for
$\sigma=1.25$ (dot-dashed), $\sigma=1$ (dashed), $\sigma=0.75$ (dotted) and 
$\sigma=0.65$ (full line). It is seen that the small particles ($\sigma<1$)
accumulate in the interfacial region. Note also that $\eta_+(0.65)>\eta_-(0.65)
$ whereas $\eta_+(0.75)<\eta_-(0.75)$ in agreement with the reversal seen in 
Fig. 2. In (c) we show $\eta(z,\sigma=0.65)$ for $\alpha=15$, $\eta_0=0.4842$ and $t=0.85$ (full line), $1$ (dots), $1.1$ (short dashes), $1.2$ (long dashes) and $1.28$ (dot-dash). (Here $z^*=(z-z_G(1))/R(1)$).\par
  {\bf FIG. 4.} The adsorption $\Gamma_1(\sigma)$ of $\sigma$-particles
 relative to the zero-adsorption Gibbs dividing surface ($z_G(1)$) of the reference
 particle ($\sigma=1$) at (a) fixed density and (b) fixed temperature. Panel (a) 
corresponds to $\alpha=15$, $\eta_0=0.4842$ and $t=0.90$ (full line) $1.00$ (dots), $1.10$ (short dashes), $1.20$ (long dashes), $1.28$ (dot-dash). Panel (b) 
corresponds to $\alpha=50$, $t=1.15$ and $\eta_0=0.45$ (line), $0.3659$ (dots),
 $0.3$ (dashes). Note the rapid variations with $t$ and $\eta_0$ of the
 interfacial excess of the small particles ($\sigma<1$) and depletion of the
 large particles ($\sigma>1$). (Here $\Gamma_1^*(\sigma)=\Gamma_1(\sigma).\, v(1)/R(1)$.)\par
 {\bf FIG. 5.} The pressure profile ($p(z)$) versus the distance ($z$) from a planar interface perpendicular to $z$-axis for (a) $\alpha=50$ and 
$\eta_0=0.3659$ and (b) $\alpha=15$ and $\eta_0=0.4842$ and three temperatures 
$t=0.9$ (full line), $t=1$ (dots) and $t=1.1$ (dot-dashes). The interfacial
 region is seen to be broadened by the polydispersity. All profiles exhibit a
pressure depletion (excess) on the high (low) density side of the interface. (Here $p^*(z)=p(z)\, v(1)/\epsilon(1,1)$ and $z^*=z/R(1)$.)\par
{\bf FIG. 6.} The surface tension ($\gamma$) versus the 
temperature ($t$) for : (a) $\alpha=50$ and $\eta_0=0.3$ (dots), $\eta_0=\eta_c=0.3659$ (full line) and $\eta_0=0.45$ (dot-dash); (b) $\alpha=15$ (dot-dash), $\alpha=50$ (dots) and $\alpha=\infty$ (full line) for $\eta_0=\eta_c(\alpha)$; (c) the same as (b) but plotted now versus $t/t_c(\alpha)$ where $t_c(\alpha)$ and $\eta_c(\alpha)$ are, respectively, the (reduced) critical-point temperature and density of a system with polydispersity index $I=1+(1/\alpha)$. 
(Here $\gamma^*=\gamma \, v(1)/\epsilon(1,1)) \, R(1)$.)\par
{\bf FIG. 7.} Tolman's length ($l_T=z_G(1)-z_0$)
 versus the temperature ($t$) for $\alpha=50$ and $\eta_0=0.3$ (dots), $\eta_0=\eta_c=0.3659$ (dash) and $\eta_0=0.45$ (dot-dash). (Here $l_T^*=l_T/R(1)$.)\par
{\bf FIG. 8.} A $ln \,\gamma^*$ versus $ln\,t^*$ plot for $\alpha=\infty$
(full line) and $\alpha=15$, $\eta_0=\eta_c=0.4842$ (dot-dash).
In both cases the critical exponent (=$3/2$) of $\gamma$ is classical.
(Here $t^*=(t_c(\alpha)-t)/t_c(\alpha)$.)
\par

\end{multicols}
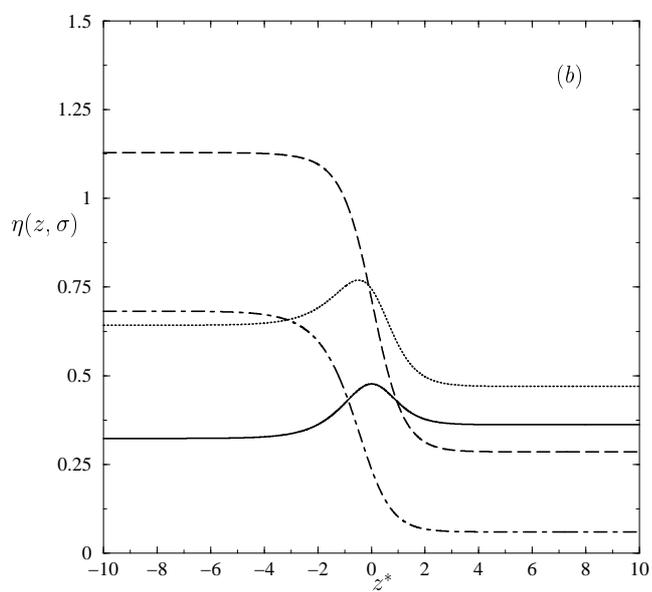
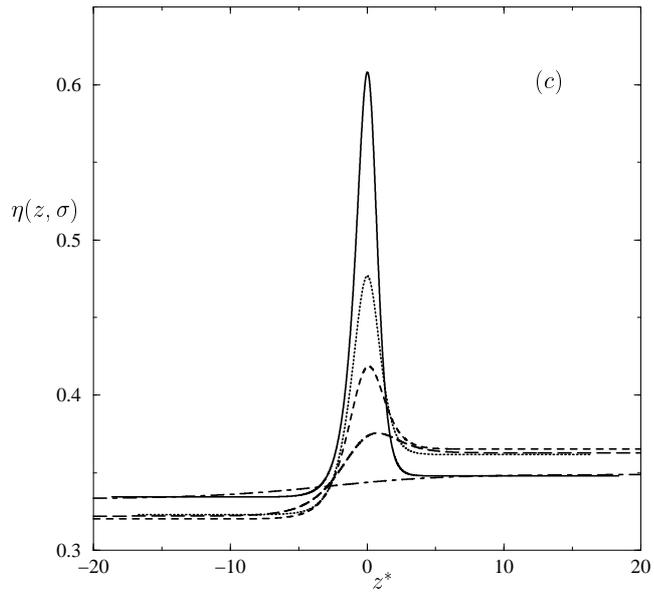
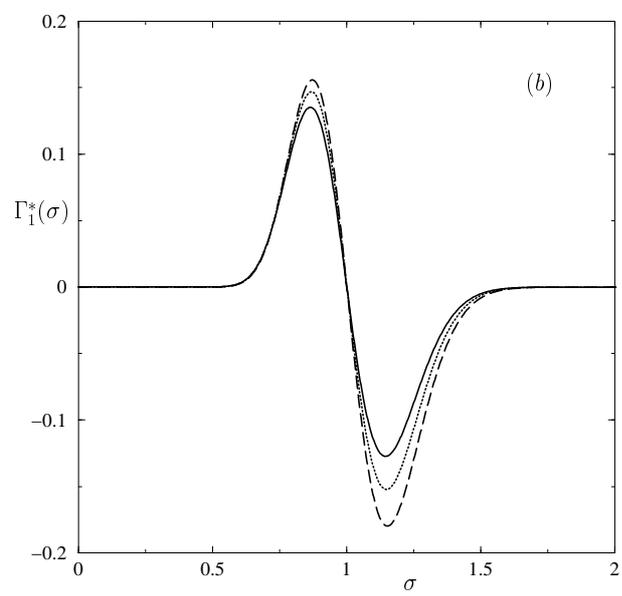
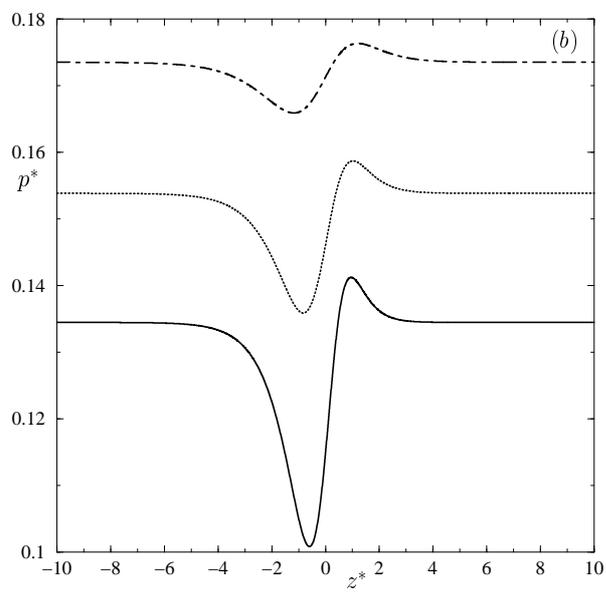
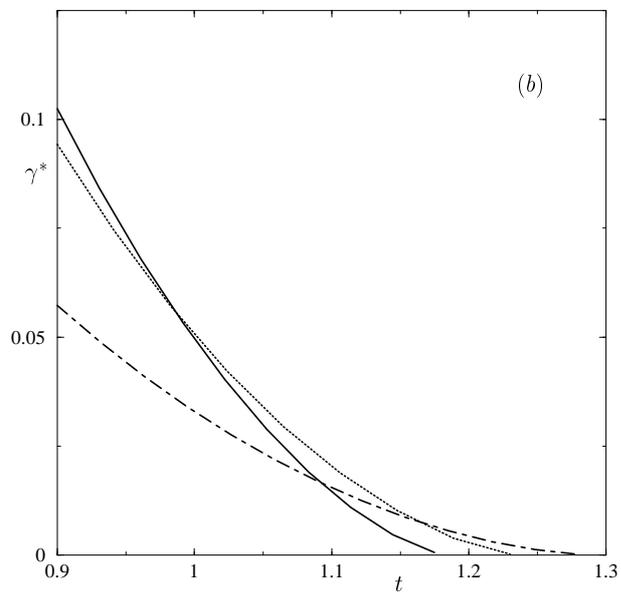
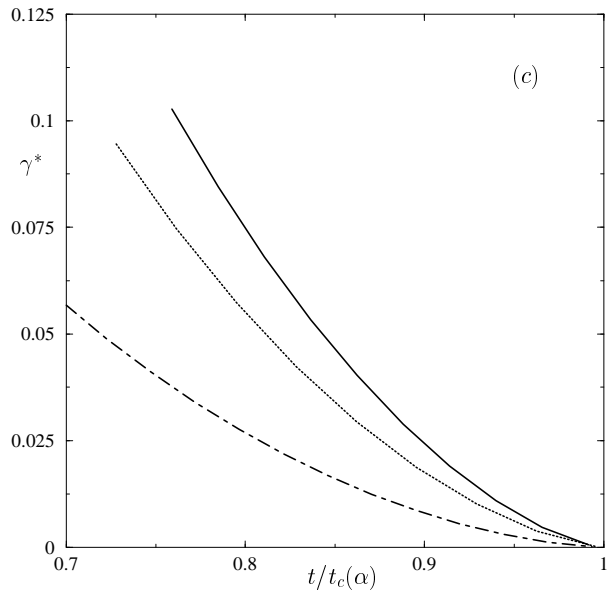
\end{document}